\documentclass[useAMS,usenatbib]{mn2e}
\usepackage{graphicx}
\usepackage{epsfig}
\usepackage{epsf}
\usepackage{amsmath}
\usepackage{amssymb}
\usepackage{stfloats}
\title[SN\,Hunt\,248]{SN\,Hunt\,248: a super-Eddington outburst from a massive cool hypergiant}

\author[Mauerhan et al.]{Jon C. Mauerhan$^{1}$\thanks{E-mail:
    mauerhan@astro.berkeley.edu}, Schuyler D. Van Dyk$^{2}$, Melissa L. Graham$^{1}$, WeiKang Zheng$^{1}$,  \newauthor
     Kelsey I. Clubb$^{1}$, Alexei V. Filippenko$^{1}$, Stefano Valenti$^{3,4}$, Peter Brown$^{5}$, \newauthor Nathan Smith$^{6}$, D. Andrew Howell$^{3,4}$, Iair Arcavi$^{3,4}$ \\
  $^{1}$Department of Astronomy, University of California, Berkeley, CA 94720-3411, USA \\ 
  $^{2}$Infrared Processing and Analysis Center, California Institute of Technology, Pasadena, CA, 91125, USA \\
  $^{3}$Department of Physics, Broida Hall, University of California, Santa Barbara, CA 93106, USA \\
  $^{4}$Las Cumbres Observatory Global Telescope Network Inc., Goleta, CA 93117 \\
  $^{5}$George P. and Cynthia Woods Mitchell Institute for Fundamental Physics \& Astronomy, Texas A\&M University, College Station, TX 77843, USA \\
$^{6}$Steward Observatory, University of Arizona, Tucson, Arizona 85721, USA}            

\begin{document}
\pagerange{\pageref{firstpage}--\pageref{lastpage}} \pubyear{2012}

\maketitle

\label{firstpage}

\begin{abstract}
We present observations of SN\,Hunt\,248, a new supernova (SN) impostor in NGC 5806, which began a multi-stage outburst in 2014 May. The `2014a' discovery brightening exhibited an absolute magnitude of $M\approx-12$ and the spectral characteristics of a cool, dense outflow, including P-Cygni lines of Fe~{\sc ii}, H~{\sc i}, and Na~{\sc i}, and line blanketing from metals. The source rapidly climbed and peaked at $M\approx-15$\,mag after two additional weeks. During this bright `2014b' phase the spectrum became dominated by Balmer emission and a stronger blue continuum, similar to the SN impostor SN\,1997bs. Archival images from the \textit{Hubble Space Telescope} between 1997 and 2005 reveal a luminous ($4\times10^{5}\,{\rm L}_{\odot}$) variable precursor star. Its location on the Hertzsprung-Russell diagram is consistent with a massive ($M_{\rm init}\approx 30\,{\rm M}_{\odot}$) cool hypergiant having an extremely dense wind and an Eddington ratio ($\Gamma$) just below unity. At the onset of the 2014a brightening, however, the object was super-Eddington ($\Gamma=4$--12). The subsequent boost in luminosity during the 2014b phase probably resulted from circumstellar interaction. SN\,Hunt\,248 provides the first  case of a cool hypergiant undergoing a giant eruption reminiscent of outbursts from luminous blue variable stars (LBVs). This lends support to the hypothesis that some cool hypergiants, such as $\rho$~Cas, could be LBVs masquerading under a pseudo-photosphere created by their extremely dense winds. Moreover, SN\,Hunt\,248 demonstrates that eruptions stemming from such stars can rival in peak luminosity the giant outbursts of much more massive systems like $\eta$~Car.
 \end{abstract}

\begin{keywords}
  circumstellar matter --- stars: evolution --- stars: winds, outflows
  --- supernovae: general --- supernovae: individual (SN\,Hunt\,248)
\end{keywords}

\section{Introduction}
In the course of hunting the skies for supernovae (SNe), modern surveys have revealed a growing sample of transients that peak at significantly fainter luminosities than normal SNe and exhibit outflow velocities well below the $\sim$10$^4$\,km\,s$^{-1}$ speeds typically observed. These so-called ``SN impostors'' (Van Dyk et al. 2000; Smith et al. 2011) are generally thought to be extragalactic analogs of the historic nonterminal eruptions of the luminous blue variable stars (LBVs) $\eta$ Carinae and P Cygni. Some well-known extragalactic examples include SN\,1954J, SN\,1997bs, and SN\,1999bw (Kochanek et al. 2012). The mass loss associated with these nonterminal eruptions can be very substantial, ranging from $0.01\,{\rm M}_{\odot}$ to $\gtrsim 10\,{\rm M}_{\odot}$ (Smith \& Owocki 2006), and it can dramatically punctuate the more persistent loss of H-rich matter driven by relatively steady stellar winds. The physical processes that trigger and drive the large outbursts remain unexplained, yet are undoubtedly a critical component of the post-main-sequence evolution of massive stars, providing a potential pathway to the Wolf-Rayet phase (see Smith 2014 for a review; but also see Smith \& Tombleson 2014). 

Observations of SN impostors have revealed a broad range of spectral morphologies, peak luminosities, and outburst timescales (Smith et al. 2011). This suggests that there could be multiple physical channels for generating giant outbursts. Possibilities include a massive star exceeding its Eddington luminosity (Owocki \& Shaviv 2012), turbulent energy transport during the late stages of core and shell burning (Smith \& Arnett 2014), or violent binary interactions (Smith 2011). Diversity in SN impostors could also result from the varying degrees of dense circumstellar matter (CSM) ejected by the stellar precursor before a major outburst. Interaction between the CSM and the relatively fast ensuing outflow can dramatically affect the luminosity, duration, and spectral characteristics of the event, operating much like a scaled-down SN\,IIn (e.g., Smith 2013; see Filippenko 1997 for a review of SN spectral classification). All of the physical processes mentioned above can presumably operate in massive stars over a broad range of initial masses, which implies that SN impostors could stem from a wide variety of stellar types beyond classical LBVs such as $\eta$ Car and P Cygni. 

To date, there are only a few direct detections of the stellar precursors to extragalactic SN impostors. By far the most well-studied example is SN\,2009ip. The precursor was a hot and extremely luminous star with $L=10^{5.9}\,{\rm L}_{\odot}$ and 50--80\,${{\rm M}_{\odot}}$ (Smith et al. 2010; Foley et al. 2011). After a decade of documented variability and multiple LBV outbursts between 2009 and 2011, the object underwent an unprecedentedly bright and energetic event in 2012, likely to be a terminal SN explosion (Mauerhan et al. 2013a, 2014; Smith, Mauerhan, \& Prieto 2014; however, also see Pastorello et al. 2013; Fraser et al. 2013; Margutti et al. 2014). The stellar parameters of the progenitor of SN\,2009ip place its effective initial mass well above the upper mass limit for red supergiants, so the object was likely to be a more luminous but also more compact blue supergiant, which is also consistent with the parameters of its purported SN light curve in 2012 (see Smith, Mauerhan, \& Prieto 2014). The source UGC~2773-OT is another example of a SN impostor with an observed stellar precursor (Smith et al. 2010; Foley et al. 2011). The star had a luminosity of $L=10^{5.1}\,{\rm L}_{\odot}$ and estimated initial mass of $\ga 20$\,${\rm M}_{\odot}$. The observed colour of the star was similar to that of an early A-type supergiant, which is a spectral type observed for some lower-luminosity LBVs. However, the object also had a dusty circumstellar environment revealed by its infrared excess, so the star could have been hotter and bluer than the source colours suggested at face value. 

Other examples of SN impostors which have documented observations of pre-outburst variability are SN\,1954J, SN\,1961V, SN\,2000ch, and the peculiar LBV + Wolf-Rayet binary HD\,5980 (Smith et al. 2011; Van Dyk \& Matheson 2012). The case of SN\,1961V, however, remains controversial; it is unclear whether the object exploded as a true SN (Smith et al. 2011; Kochanek et al. 2011). SN\,1997bs is another example, for which there is a single pre-outburst observation of a luminous supergiant star at the position of the event (Van Dyk et al. 2000). 

Transients with pre-outburst detections of relatively low-mass progenitors that are apparently not consistent with classical LBV phenomena include NGC\,300-OT (Prieto et al. 2009; Thompson et al. 2009; Berger et al. 2009) and SN\,2008S (Prieto et al. 2008), both of which had a luminous mid-infrared counterpart indicative of a heavily dust-enshrouded object, perhaps a super asymptotic-giant-branch (AGB) star. SN\,2010U and M85-OT are other luminous outbursts that potentially arose from a relatively low-mass stars, although the uncertain influence of extinction associated with these objects has complicated the characterisation of their stellar precursors.  

Here we present observations of a new SN impostor, SN\,Hunt\,248, which experienced a luminous multi-stage outburst beginning May 2014. This object is particularly interesting because the precursor star was detected on multiple occasions in space-based archival images with the \textit{Hubble Space Telescope (HST)}, and the multi-colour data indicate that the star was in a cool hypergiant phase. SN\,Hunt\,248 thus provides the first direct connection between cool hypergiants and eruptive LBV phenomena, a result which has important implications for the evolution of massive stars. 

\begin{figure}
\includegraphics[width=3.3in]{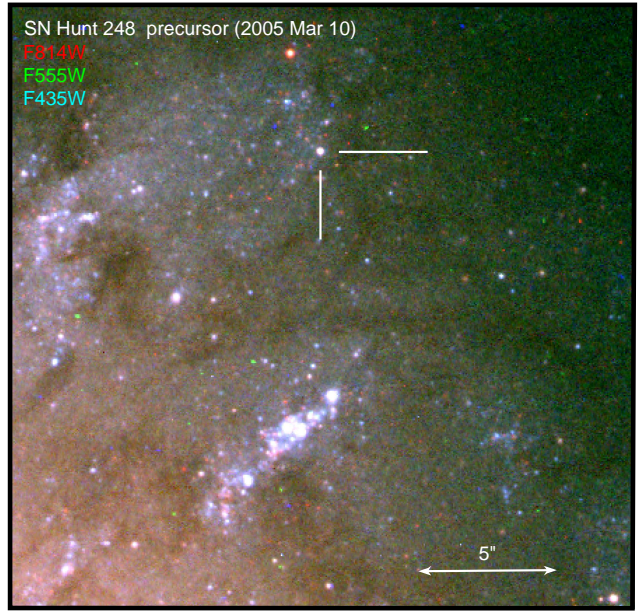}
\caption{{\it HST} colour-composite image showing the SN\,Hunt\,248 precursor in NGC~5806 on 2005 Mar. 10. $RGB$ colours correspond to images in F435W, F606W, and F814W, respectively. North is up and east to the left.}
\label{fig:spec_wide}
\end{figure}

\section{Observations and Results}

\begin{table*}\tiny\begin{center}\begin{minipage}[bp]{3.2in} \setlength{\tabcolsep}{2.pt}
      \caption{KAIT, Nickel, and LCOGT Photometry of SN\,Hunt\,248.}
\centering
  \begin{tabular}[bp]{@{}lcccccc@{}} 
  \hline
KAIT, Nickel \\
 2014 UT Date    & $B$  & $V$ & $R$  & $I$    & clear \\                            
    (JD--2,450,000)       & (mag) & (mag) & (mag) & (mag) & (mag) \\
\hline
\hline
5-21 (6798.7)$^a$      &                 $\dots$                  &      $\dots$                &             $\dots$                   &            $\dots$      &      $\sim20.3^{a}$   \\
5-22 (6799.8)$^b$      &                 $\dots$                  &      $\dots$                &             $\dots$                   &            $\dots$      &      $20.04\pm0.46$$^{b}$   \\
5-29 (6807.3)$^c$        &                 $\dots$                  &      $\dots$                &             $\dots$                   &            $\dots$      &      $\sim19.8^{c}$   \\
6-03 (6812.35)      &                 $\dots$                  &     $\dots$                 &             $\dots$                   &           $\dots$       &     $19.66\pm0.27$     \\
6-04 (6813.33)      &                 $\dots$                  &    $\dots$                  &              $\dots$                   &         $\dots$        &     $18.75\pm0.21$\\
6-05 (6814.32)      &                 $\dots$                  &    $\dots$                  &               $\dots$                  &         $\dots$        &     $18.24\pm0.10$\\
6-06 (6815.36)      &                $\dots$ &    $\dots$ &               $\dots$                  &         $17.93\pm0.21$            &     $17.94\pm0.07$\\
6-07 (6816.22)      &                 $18.77\pm0.29$ &    $18.18\pm0.16$ &               $17.88\pm0.09$                  &         $17.73\pm0.12$            &     $17.88\pm0.06$\\
6-08 (6817.25)      &                 $18.48\pm0.32$ &    $18.05\pm0.11$ &               $17.72\pm0.09$                  &         $17.52\pm0.12$            &     $17.62\pm0.06$\\
6-11 (6820.32)      &                 $17.68\pm0.21$      &    $17.41\pm0.12$  &               $17.01\pm0.08$                  &         $17.05\pm0.11$            &     $16.97\pm0.06$\\
6-12 (6821.31)      &                 $17.90\pm0.25$               &    $17.23\pm0.15$  &               $17.04\pm0.09$                  &         $16.70\pm0.09$            &     $16.91\pm0.06$\\

6-12 (6821.41)-N       &                 $17.40\pm0.05$               &    $17.15\pm0.03$  &               $16.91\pm0.02$                  &         $16.79\pm0.02$            &     $\dots$   \\

6-13 (6822.38)      &                 $17.68\pm0.21$                 &    $17.20\pm0.12$ &               $16.78\pm0.08$                  &         $16.81\pm0.11$            &     $16.84\pm0.06$\\
6-14 (6823.31)      &                 $17.58\pm0.23$                  &    $17.08\pm0.12$  &               $16.87\pm0.06$                  &         $16.78\pm0.08$            &     $16.82\pm0.06$\\
6-15 (6824.28)      &                 $17.33\pm0.12$ &    $17.11\pm0.08$  &               $16.79\pm0.05$                  &         $16.73\pm0.06$            &     $16.77\pm0.06$\\

6-15 (6824.41)-N       &                 $17.45\pm0.09$                  &      $\dots$                &             $\dots$                   &            $\dots$           &     $\dots$   \\

6-16 (6825.29)      &                 $17.51\pm0.15$               &      $\dots$                 &               $16.75\pm0.06$                  &         $16.72\pm0.08$            &     $16.75\pm0.05$\\
6-17 (6826.29)      &                 $17.50\pm0.09$ &    $17.10\pm0.04$ &               $16.81\pm0.04$                  &         $16.69\pm0.06$            &     $16.79\pm0.04$\\
6-18 (6827.29)      &                 $17.45\pm0.07$ &    $17.19\pm0.05$ &               $16.79\pm0.04$                  &         $16.72\pm0.06$            &     $16.79\pm0.04$\\
6-19 (6828.28)      &                 $17.35\pm0.06$                 &    $17.12\pm0.04$&               $16.77\pm0.04$                  &         $16.69\pm0.06$            &     $16.74\pm0.03$\\
%6-20 (6829.25)      &                 $\dots$                   &    $17.13\pm0.04$  &               $16.78\pm0.06$                  &         $16.40\pm0.14$            &     $16.79\pm0.13$\\
6-21 (6830.28)      &                 $17.42\pm0.09$  &    $17.05\pm0.05$   &               $16.74\pm0.04$                  &         $16.65\pm0.05$            &     $16.74\pm0.02$\\
6-22 (6831.27)      &                 $17.48\pm0.08$  &    $17.12\pm0.04$&               $16.80\pm0.04$                  &         $16.70\pm0.05$            &     $16.78\pm0.04$\\
6-23 (6832.24)      &                 $17.44\pm0.08$  &    $17.18\pm0.06$  &               $16.84\pm0.04$                  &         $16.76\pm0.05$            &     $16.82\pm0.02$\\
6-24 (6833.27)      &                 $17.62\pm0.11$                    &    $17.25\pm0.06$&               $16.88\pm0.04$                  &         $16.79\pm0.05$            &     $16.87\pm0.04$\\
6-26 (6835.22)      &                 $17.67\pm0.08$  &    $17.32\pm0.06$  &               $17.00\pm0.05$                  &         $16.90\pm0.05$            &     $16.98\pm0.02$\\
6-27 (6836.26)      &                    $\dots$                  &      $\dots$                &             $\dots$                                     &                                   $\dots$    &     $17.01\pm0.05$\\
6-28 (6837.22)      &                 $17.85\pm0.09$  &    $17.53\pm0.05$  &               $17.17\pm0.04$                  &         $17.05\pm0.07$            &     $17.07\pm0.03$\\
6-29 (6838.35)      &                 $17.87\pm0.10$  &    $17.52\pm0.06$  &               $17.22\pm0.04$                  &         $17.12\pm0.06$            &     $17.17\pm0.04$\\

6-29 (6838.38)-N      &                 $17.83\pm0.02$  &    $17.52\pm0.02$  &               $17.16\pm0.02$                  &         $17.09\pm0.04$            &     $\dots$\\

6-30 (6839.24)      &                 $17.95\pm0.10$  &    $17.66\pm0.10$  &               $17.23\pm0.07$                  &         $17.12\pm0.10$            &     $17.10\pm0.11$\\
7-01 (6840.22)      &                 $18.15\pm0.11$  &    $17.74\pm0.06$  &               $17.36\pm0.05$                  &         $17.30\pm0.08$            &     $17.36\pm0.04$\\
7-01 (6840.29)-N      &                 $17.98\pm0.02$  &    $17.63\pm0.01$  &               $17.32\pm0.01$                  &         $17.21\pm0.02$            &     $\dots$\\

7-04 (6841.23)      &                 $18.08\pm0.14$  &    $17.75\pm0.07$  &               $17.42\pm0.05$                  &         $17.27\pm0.09$            &     $17.40\pm0.03$\\

7-05 (6842.24)      &                 $18.27\pm0.13$  &    $17.87\pm0.08$  &               $17.54\pm0.05$                  &         $17.31\pm0.07$            &     $17.48\pm0.04$\\
7-06 (6843.24)      &                 $18.37\pm0.14$  &    $17.87\pm0.07$  &               $17.58\pm0.07$                  &         $17.37\pm0.07$            &     $17.53\pm0.03$\\
7-07 (6844.22)      &                 $18.50\pm0.17$  &    $18.03\pm0.10$  &               $17.66\pm0.06$                  &         $17.37\pm0.10$            &     $17.64\pm0.04$\\
7-11 (6848.21)      &                 $\dots$  &    $17.93\pm0.14$  &               $17.75\pm0.11$                  &         $17.41\pm0.12$            &     $17.90\pm0.06$\\
7-12 (6849.22)      &                 $\dots$  &    $18.47\pm0.25$  &               $17.82\pm0.12$                  &         $17.58\pm0.15$            &     $17.82\pm0.07$\\
7-12 (6849.29)-N      &                 $18.70\pm0.09$  &               $18.15\pm0.05$                  &         $17.84\pm0.03$            &     $17.71\pm0.03$ & $\dots$\\

\hline
\hline
LCOGT & $B$  & $V$ & $g^{\prime}$  & $r^{\prime}$ & $i^{\prime}$ \\
\hline
6-06 (6815.3)      &                 $18.54\pm0.08$ &    $18.53\pm0.07$  &              $18.44\pm0.04$              &         $18.22\pm0.07$         &     $18.05\pm0.10$ \\
6-13 (6822.0)      &                 $17.32\pm0.02$ &    $17.21\pm0.12$  &               $17.22\pm0.02$              &         $16.97\pm0.02$            &     $16.98\pm0.02$\\
6-16 (6825.0)      &                 $17.24\pm0.01$ &      $17.03\pm0.01$&                $17.16\pm0.01$                   &          $16.91\pm0.01$           &     $16.99\pm0.01$ \\
6-18 (6827.0)  &                 $17.31\pm0.01$ &      $17.04\pm0.01$&                    $17.20\pm0.01$                   &          $16.92\pm0.01$           &     $\dots$ \\
6-19 (6828.1)   &                 $17.33\pm0.01$ &         $17.06\pm0.01$                  &         $\dots$                        &         $\dots$                          &         $\dots$             \\
6-20 (6829.2)    &              $17.30\pm0.01$ &          $17.08\pm0.01$                  &         $\dots$                        &         $\dots$                          &         $\dots$             \\
6-22 (6831.2)     &         $17.32\pm0.01$ &          $17.13\pm0.01$                  &         $\dots$                        &         $\dots$                          &         $\dots$             \\
6-23 (6832.2)&                  $17.39\pm0.01$ &          $17.14\pm0.01$                  &         $\dots$                        &         $\dots$                          &         $\dots$             \\
6-24 (6833.2) &                  $17.37\pm0.01$ &             $\dots$                &         $\dots$                        &         $\dots$                          &         $\dots$             \\
6-26 (6835.1)&                 $17.57\pm0.01$ &    $17.28\pm0.01$  &                   $\dots$                        &         $\dots$                          &         $\dots$             \\
6-27 (6836.2)&                 $17.66\pm0.01$ &    $17.34\pm0.01$  &              $17.47\pm0.01$              &         $17.12\pm0.01$         &     $17.21\pm0.01$ \\
6-28 (6837.0)&                 $17.73\pm0.01$ &    $17.40\pm0.01$  &              $17.58\pm0.01$              &         $\dots$                           &     $17.31\pm0.02$ \\
6-30 (6839.0) &                 $17.87\pm0.01$ &    $17.57\pm0.01$  &              $17.71\pm0.01$              &         $17.35\pm0.01$         &     $17.43\pm0.01$ \\
7-03 (6842.0) &                 $18.13\pm0.02$ &    $17.79\pm0.03$  &              $18.04\pm0.04$              &         $17.59\pm0.04$         &     $17.57\pm0.02$ \\
7-05 (6844.2) &                 $18.63\pm0.04$ &    $17.93\pm0.01$  &              $18.05\pm0.01$              &         $17.64\pm0.01$         &     $17.67\pm0.01$ \\
\hline
\end{tabular} 
\tiny{(a) Photometry from CRTS (S. Howerton); (b) J. Brimacombe; (c) G. Masi.}
\tiny{Dates of Nickel photometry are marked with ``-N". Uncertainties are statistical ($1\sigma$).}
\end{minipage} \end{center}
\end{table*}

\begin{table*}\tiny\begin{center}\begin{minipage}[bp]{3.2in} \setlength{\tabcolsep}{2.pt}
      \caption{\textit{Swift}/UVOT Photometry of SN\,Hunt\,248.}
\centering
  \begin{tabular}[bp]{@{}lccccccc@{}} 
  \hline
\textit{Swift} UVOT \\
 2014 UT Date    & $V$  & $B$ & $U$  & $UVW1$    & $UVM2$   & $UVW2$ \\                            
    (JD--2,450,000)       & (mag) & (mag) & (mag) & (mag) & (mag) & (mag) \\
\hline
\hline
6-07 (6816.94)            &      $18.25\pm   0.22   $        &     $18.33 \pm  0.11 $        &     $17.89 \pm  0.11  $      &      $20.08 \pm  0.37  $      &         $\dots$          &       $\dots$         \\
6-11 (6820.86)            &     $17.44 \pm  0.13    $       &      $17.55 \pm  0.09  $      &      $ 16.85  \pm 0.08 $     &     $ 18.87 \pm  0.17  $      &       $20.06 \pm  0.31 $       &     $ 19.40 \pm  0.20 $         \\
6-13 (6822.47)             &      $\dots$                        &      $ 17.32\pm   0.10 $      &    $ 16.92  \pm 0.11 $       &     $  18.37\pm   0.16 $      &     $\dots$       &            $\dots$        \\
6-15 (6824.13)             &     $17.18 \pm  0.12    $        &      $17.28 \pm  0.08  $      &     $ 16.69  \pm 0.08 $      &      $ 18.31\pm   0.13 $      &     $ 19.09\pm   0.17 $           &  $  18.89\pm   0.16 $           \\
6-17 (6826.80)             &      $17.37 \pm  0.14  $        &       $17.37 \pm  0.09$       &       $16.66\pm   0.08 $     &      $ 18.15 \pm  0.13  $     &      $ 18.90\pm   0.17 $       &     $ 18.75 \pm  0.15 $         \\
6-19 (6828.55)             &      $17.15\pm   0.11  $        &     $ 17.28  \pm 0.08 $        &       $16.56 \pm  0.08$      &     $  18.01 \pm  0.11 $       &    $ 19.05\pm   0.16 $         &    $ 18.66\pm   0.14$           \\
6-21 (6830.82)            &      $17.08\pm   0.11  $       &       $  17.31\pm   0.08 $    &         $16.57\pm   0.08 $   &      $ 17.74 \pm  0.10 $       &      $ 18.44\pm   0.12 $       &      $  18.50 \pm  0.13 $       \\
6-23 (6832.91)             &      $17.33 \pm  0.18   $      &        $17.49 \pm  0.11 $     &         $16.72\pm   0.11 $    &      $ 17.70 \pm  0.14 $      &       $\dots$          &      $ 18.28 \pm  0.15 $        \\
6-25 (6834.55)             &      $17.42 \pm  0.14  $       &          $17.59\pm   0.09 $    &         $16.88\pm   0.09$    &      $  17.91 \pm  0.12 $     &      $18.61 \pm  0.13 $        &    $  18.49\pm   0.13 $         \\
6-27 (6836.35)             &      $17.54 \pm  0.14 $       &         $ 17.67 \pm  0.09 $   &         $ 17.17 \pm  0.09 $  &        $ 18.01 \pm  0.12 $    &       $ 18.62 \pm  0.13  $     &    $ 18.90 \pm  0.15 $          \\
7-01 (6840.12)            &      $17.81 \pm  0.17$        &         $18.08  \pm 0.11 $     &         $17.62 \pm  0.11 $    &      $ 18.35 \pm  0.13  $     &       $  19.16 \pm  0.17  $    &     $ 19.35  \pm 0.20$          \\
7-03 (6842.95)            &      $18.01 \pm  0.19$        &         $18.35  \pm 0.13 $     &         $17.98 \pm  0.14 $    &      $ 18.48 \pm  0.15  $     &       $  19.04 \pm  0.16  $    &     $ 19.65  \pm 0.24$          \\
7-05 (6844.55)            &      $17.83 \pm  0.22$        &         $18.64  \pm 0.15 $     &         $18.51 \pm  0.18 $    &      $ 18.80 \pm  0.17  $     &       $  19.31 \pm  0.24  $    &     $ 19.39  \pm 0.22$          \\
7-08 (6847.86)            &      $18.39 \pm  0.33$        &         $18.59  \pm 0.19 $     &         $18.65 \pm  0.25 $    &      $ 19.08 \pm  0.24  $     &       $  19.17 \pm  0.30  $    &     $\dots$          \\
7-09 (6848.34)            &      $18.40 \pm  0.33$        &         $18.64  \pm 0.16 $     &         $18.60 \pm  0.20 $    &      $ 18.99 \pm  0.20  $     &       $\dots $                            &     $19.94\pm0.29$          \\
\hline
\end{tabular}
\tiny{Uncertainties are statistical ($1\sigma$).}
\end{minipage} \end{center}
\end{table*}

\begin{table*}\tiny\begin{center}\begin{minipage}[bp]{3.2in} \setlength{\tabcolsep}{2.pt}
      \caption{{\it HST} and SDSS photometry of the SN\,Hunt\,248 precursor.}
\centering
  \begin{tabular}[bp]{@{}lccccr@{}} 
  \hline
 Date      & $B$  & $V$ & $I$ or $i$  & F658N  & Obs. ID, PI\\                            
 (JD--2,450,000)          & (mag) & (mag) & (mag) & (mag) & \\
\hline
\hline
1997-04-30 (569)        &                 $\dots$               &  $22.70\pm0.02$  &             $\dots$                   &            $\dots$ &   6359, Stiavelli   \\ % (0568.6) 
2000-05-09 (1669)           &                 $\dots$               &  $\dots$  &                   $21.93\pm0.19$ &      $\dots$                   &   SDSS   \\
2001-07-05 (2095)      &  $21.63\pm0.03$  &             $\dots$                      & $20.96\pm0.03$ &         $\dots$  &   9042, Smartt              \\ % (6175.6)   
2004-04-03 (3098)   &              $\dots$                     &            $\dots$                       &  $22.10\pm0.01$& $20.98\pm0.03$ & 9788, Ho \\ % (6178.1)     
2005-03-10  (3439)      &  $23.30\pm0.02$  & $22.91\pm0.01$     & $22.10\pm0.01$  &          $\dots$  &    10187, Smartt                 \\ %  (6180.8) 
\hline
\end{tabular} 
\tiny{The {\it HST} flight-system magnitudes were transformed to $BVI$ following Holtzman et al. (1995) and Sirianni et al. (2005). The SDSS photometry is in the $i$ band, and was extracted via PSF-fitting photometry (see text). The uncertainties are statistical ($1\sigma$).}
\end{minipage} \end{center}
\end{table*}

\subsection{Photometry}
SN\,Hunt\,248, also known as PSN J14595947+0154262, was discovered in outburst on 2014 May 21.20 (UT dates are used throughout this paper) by Stan Howerton in the Catalina Real-Time Transient Survey (CRTS). A second, brighter outburst from this object was detected a few weeks later, prompting us to alert the community (Zheng et al. 2014).

We began photometrically monitoring the 2014 activity of SN\,Hunt\,248 on June 4 using the the 0.76\,m Katzman Automatic Imaging Telescope (KAIT; Filippenko et al. 2001) at Lick Observatory, roughly 2 weeks after its May 21 discovery. Images were obtained unfiltered and through $BVRI$ filters, and photometry was extracted via an automated reduction pipeline. $BVRI$ photometry was also obtained using the Nickel 1\,m telescope at Lick.  Calibration of the KAIT and Nickel photometry was performed using three bright field stars in the same image as SN\,Hunt\,248, for which we assigned magnitudes from the Las Cumbres Observatory Global Telescope Network (LCOGT.net; Brown et al. 2013) observations of the transient iPTF13bvn in the same host galaxy (Cao et al. 2013).  Table~1 lists the KAIT and Nickel photometry, presented in Vega magnitudes. 

Additional photometric monitoring was also initiated with LCOGT including the 1\,m facilities at Cerro Tololo (Chile), McDonald Observatory (USA), and Sutherland (South Africa). Images were obtained through $BVg^{\prime}r^{\prime}i^{\prime}$ filters and photometry was extracted via an automated software pipeline. More field stars were available for photometric calibration of the LCOGT data than for the KAIT data. Table~1 includes the LCOGT photometry. 

Space-based photometry was obtained using the \textit{Swift} satellite and UVOT instrument (Gehrels et al. 2004). Images were obtained through $UBV$ filters, and template images of the host galaxy were subtracted from the data prior to photometric extraction. Details of the reduction methods are described by Brown et al. (2014), and the resulting photometry is listed in Table~2.

For some epochs there is a $\sim0.2$--0.3\,mag discrepancy between the photometry from KAIT and that of Nickel, LCOGT, and \textit{Swift}/UVOT, particularly for the KAIT $B$-band filter. Though partly caused by differences in the number of calibration stars, this is mainly the result of the fact that the KAIT $B$ filter has a redder transmission function, with a central wavelength ($\lambda_c$) of 4445\,{\AA} and a full width at half-maximum intensity (FWHM) of 907\,{\AA} (Ganeshalingam et al. 2010). The LCOGT and UVOT $b$ filters have ($\lambda_c$, FWHM) = (4261, 887\,{\AA}) and (4392, 975\,{\AA}), respectively (Poole et al. 2008; LCOGT.net). The magnitude of the resulting photometric discrepancy depends on the spectral energy distribution of the source within the passband, and can thus change as the continuum and emission-line spectrum of SN\,Hunt\,248 evolve.

We also extracted photometry from \textit{HST} Wide-Field and Planetary Camera 2 (WFPC2) and Advanced Camera for Surveys (ACS) archival images, shown in Figure~1. The host galaxy, NGC 5806, was observed on dates between 1997 Apr. 30 and 2005 Mar. 10 through the F435W, F450W, F555W, F606W, F814W, and F658N filters. Table 3 lists the photometry extracted from these images using {\sc dolphot} (Dolphin 2000a,b). {\sc dolphot} automatically transforms the {\it HST} flight-system magnitudes to Johnson-Cousins $BVI$, following Holtzman et al. (1995) for WFPC2 and Sirianni et al. (2005) for ACS. To verify spatial coincidence of the outbursting source with the stellar precursor identified in the {\it HST} images, we used an $R$-band Nickel image obtained at Lick  on 2014 July 1 as an astrometric reference. We found 12 stars in common between this image and the {\it HST} F814W ACS mosaic. The position of the recent transient eruption in the earlier {\it HST} image, based on the Nickel image astrometry, is an ellipse with ($x,~y$) axes of (3.38, 2.61) ACS/WFC pixels (i.e., 0\farcs17, 0\farcs13), which coincides, to within the uncertainties, with the position of the stellar precursor. In addition, using 20 stars as fiducial sources and the USNO-B1 catalog as reference, we measure an absolute source position of $\alpha=14^{\rm h}59^{\rm m}59.47^{\rm s}$, $\delta = +1^{\circ}54'$26\farcs6, $\pm 0$\farcs24 (rms) in each coordinate, which differs by $0\farcs4$ in $\delta$ from the CRTS discovery position. 

The stellar precursor was also detected in $i$-band images from the Sloan Digital Sky Survey (SDSS), which were obtained on 2000 May 4. We extracted photometry using the program {\sc starfinder} written in {\sc interactive\,data\,language\,(idl)}. The software fits a model point-spread function (PSF) to stars automatically detected in the image and constructs a two-dimensional model of the background emission, which effectively accounts for the low-spatial-frequency emission from the host galaxy. We supplied {\sc starfinder} with a PSF model constructed from data obtained online\footnote{das.sdss.org/www/html}. The model PSF was constructed specifically for the section of the $i$ band containing SN\,Hunt\,248, using the {\sc idl} program {\sc sdss\_psfrec.pro}\footnote{https://code.google.com/p/sdssidl/}. Using 9 nearby field stars as a photometric reference, we measure an $i$-band magnitude of 21.93$\pm$0.19. There is a possible source discernible in the SDSS $r$ and $g$ images as well, although this is possibly a confused blend of background sources. Because of this more complex and mottled background emission from the host in the $g$ and $r$ images, we were unable to extract reliable photometry or meaningful limits for these wavelengths.  

\begin{figure}
\includegraphics[width=3.3in]{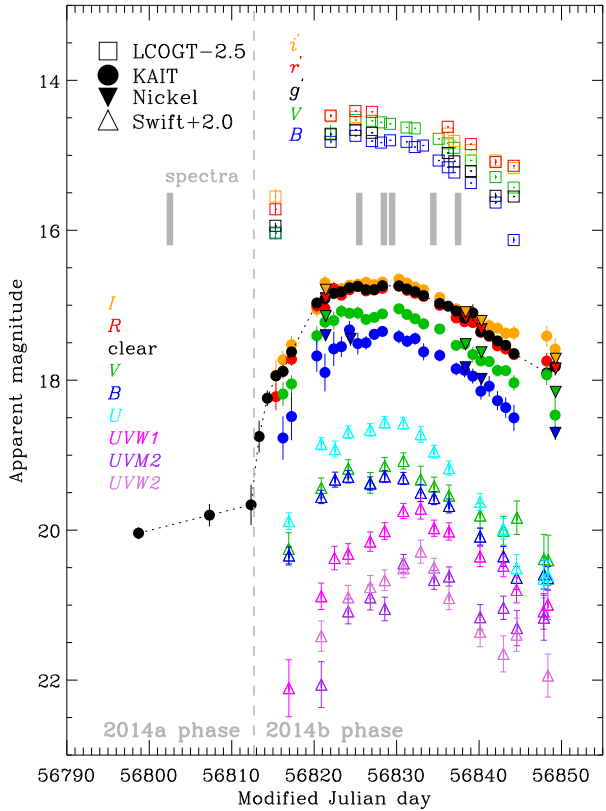}
\caption{KAIT, Nickel, LCOGT, and \textit{Swift} photometry of SN\,Hunt\,248. The earliest two unfiltered measurements are from IAU CBAT transient follow-up reports (J. Brimacombe, G. Masi), and are part of the ``2014a'' phase. The subsequent brightening is referred to as the ``2014b'' phase. The vertical grey lines mark our spectroscopic epochs.}
\label{fig:kait}
\end{figure}

\begin{figure}
\includegraphics[width=3.3in]{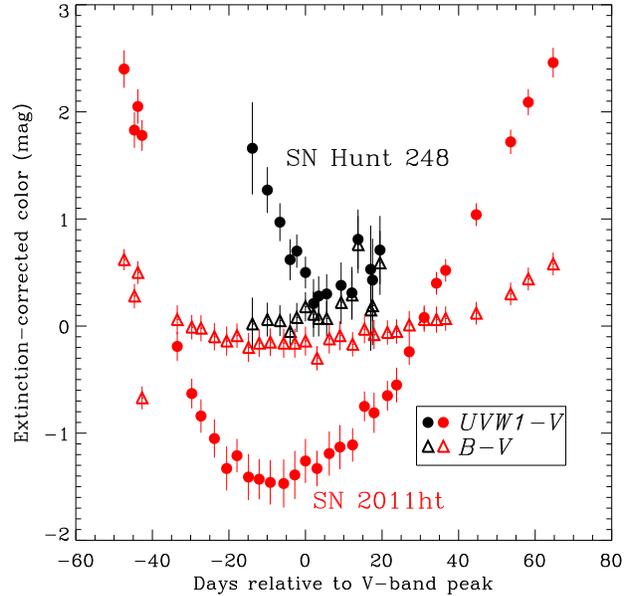}
\caption{\textit{Swift}/UVOT colour evolution of SN\,Hunt\,248 (black) compared with the Type IIn SN 2011ht (red; Roming et al. 2012). Extinction correction was applied using $A_V=0.14$ and 0.19\,mag for SN\,Hunt\,248 and SN\,2011ht, respectively, and translated to the $B$ and $UVW1$ bands assuming $R_V=3.1$. SN\,2011ht exhibits a much bluer $UVW1-V$ colour than SN\,Hunt\,248, while the $B-V$ colours of both objects are more comparable and have a relatively flat evolution. The comparison underscores the significant energetic difference between these two events.}
\label{fig:color}
\end{figure}

The distance modulus of NGC~5806 is $32.11\pm0.15$\,mag, calculated by averaging all of the individual distance measurements within the NASA Extragalactic Database (NED)\footnote{http://ned.ipac.caltech.edu/cgi-bin/nDistance?name=NGC+5806} made using the Tully-Ficher relation. This corresponds to a physical distance of $26.4\pm1.8$\,Mpc. The Galactic extinction toward NGC~5806 is estimated to be $A_V=0.14$\,mag (Schlafly \& Finkbeiner 2011). Absolute magnitudes were derived using these values.

\subsubsection{Light curve of the precursor star and outburst}
\begin{figure*}
\includegraphics[width=7.in]{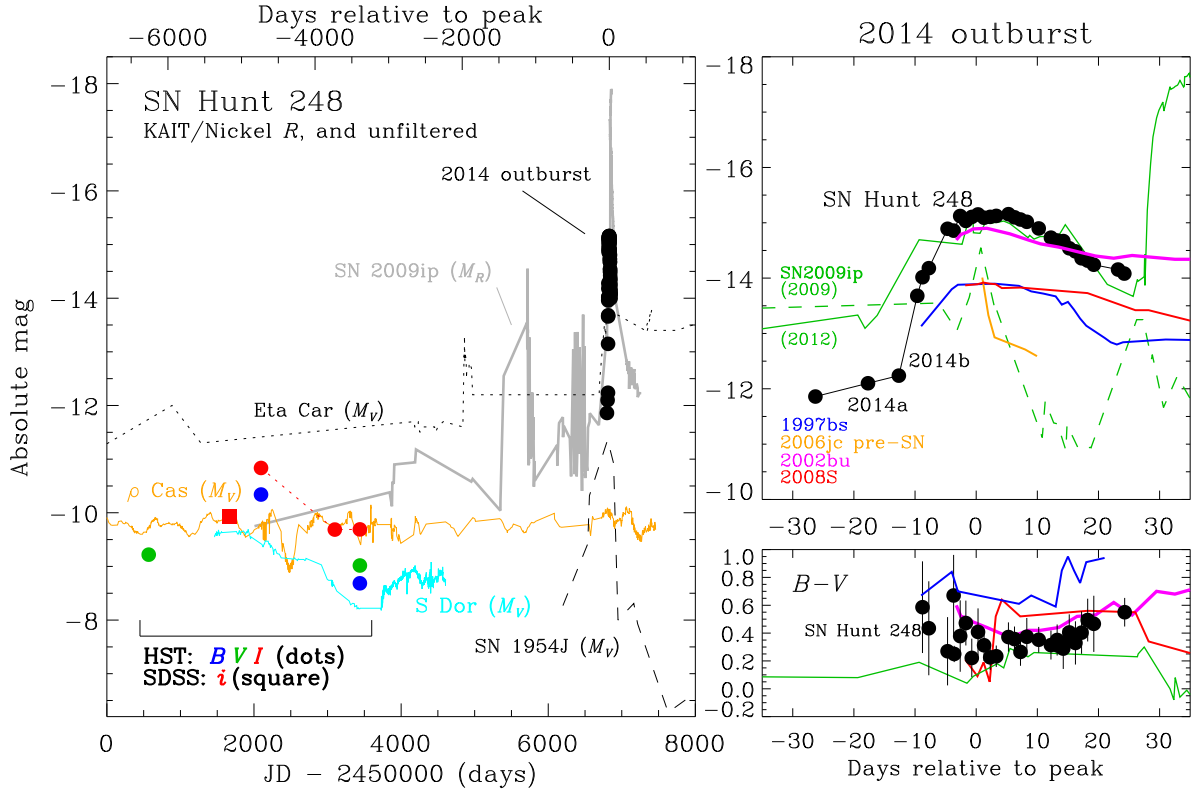}
\caption{\textit{Left panel:} Long-term light curve of SN\,Hunt\,248, including {\it HST} $BVI$ and SDSS $i$ photometry of the precursor between 1997 April and 2005 March (coloured filled circles and square, respectively), and recent photometry of its luminous outburst during 2014 (KAIT/Nickel, black filled circles). Absolute magnitudes for the outburst photometry were derived from the unfiltered (for the 2014a phase) and $R$-band measurements (which closely match unfiltered values), and no bolometric correction was applied. Extinction correction was performed using $A_V=0.14$\,mag applied to other bands assuming $R_V=3.1$. The historic light curves of $\eta$~Car (Smith \& Frew 2011), SN\,1954J (Tammann \& Sandage 1968), and SN\,2009ip (Pastorello et al. 2013; Graham et al. 2014) are plotted as dotted, dashed, and solid grey curves, respectively. \textit{Right panels}: Detail of the 2014 outburst light curve and observed $B-V$ colour evolution (no extinction correction) of SN\,Hunt\,248, compared with other SN impostors, including SN\,1997bs (Van Dyk et al. 2000), SN\,2006jc (Pastorello et al. 2007), SN\,2002bu (Smith et al. 2011), SN\,2009ip (Mauerhan et al. 2013a), and SN\,2008S (Smith et al. 2009). A portion of the historic light curves of the cool hypergiant $\rho$ Cas (solid orange curve; AAVSO) and the iconic LBV S Doradus (cyan curve; AAVSO), shifted to arbitrary times, are also shown for comparison. Their respective absolute $V$ magnitudes were derived using $d = 2.5$ kpc (Humphreys 1978) and $A_V=2.3$\,mag (Yamamuro et al. 2007), and $d = 49$ kpc and $A_V=0.12$\,mag (Massey 2000). }
\label{fig:lc}
\end{figure*}

The KAIT, Nickel, LCOGT, and \textit{Swift}/UVOT light curves of apparent photometry are shown in Figure~\ref{fig:kait}, which also includes two early-time unfiltered measurements listed from the Central Bureau of Astronomical Telegrams (CBAT; photometry by J. Brimacombe and G. Masi). The $B-V$ and $UVW1-V$ colour curves are shown in Figure~\ref{fig:color}. The complete absolute magnitude light curve of SN\,Hunt\,248, including the {\it HST} photometry of the precursor, is displayed in Figure~\ref{fig:lc}, plotted along with several comparison objects  for reference. The photometry from 1997 through 2005 reveals a luminous object at the location of the transient. The source exhibits a minimum brightness of 22.9\,mag in the $V$ band, before and after a $>1$\,mag brightness increase during July 2001. Two consecutive $I$-band measurements obtained $\sim1$\,yr apart, on 2004 Apr. 3 and 2005 Mar. 10, exhibit nearly identical values. It thus seems plausible that the {\it HST} photometry from 2005 represents a state of relative ``quiescence'' for the star.  With our adopted distance modulus and extinction value discussed above, the quiescent and bright precursor phases correspond to absolute magnitudes of $M_V \approx -9$ and $M_V \approx -10.5$, respectively. 

Nearly 10\,yr after the most recent serendipitous detection of the precursor by {\it HST}, the source was discovered in outburst on 2014 May 21 at $20.04\pm0.46$\,mag (CBAT; J. Brimacombe); we refer to this as the ``2014a'' phase. Between May 21 and June 3 the source slowly increased its brightness by $<1$\,mag. Then, on June 4, it began to brighten rapidly, and by June 16 it peaked at $R=16.7$\,mag. We refer to this as the ``2014b'' phase. After peaking, the object  plateaued for $\sim 10$ days in the optical before beginning to fade significantly. The \textit{Swift} ultraviolet (UV) bands during the 2014b phase peak $\sim 5$ days later than the optical. 

The absolute magnitude of SN\,Hunt\,248 at the onset of the 2014a phase was $\sim -12$, corresponding to a luminosity of $2\times10^{40}$\,erg\,s$^{-1}$  or $L=5\times10^{6}\,{L}_{\odot}$ (derived from unfiltered photometry, which generally matches the $R$-band values closely, and using $A_R=0.11$\,mag, with no bolometric correction). The peak of the brighter 2014b phase reached $\sim -15$\,mag, which corresponds to a luminosity of $3\times10^{41}$\,erg\,s$^{-1}$ ($L\approx8\times10^{7}\,{\rm L}_{\odot}$). The peak level during 2014b is within $\sim 1$\,mag of the peak outburst levels of SN\,2009ip's discovery eruption in 2009 (Figure~\ref{fig:lc}, upper-right panel, green dashed curve). However, the outburst timescale of SN\,Hunt\,248 is not nearly as rapid as the 2009--2011 eruptions of SN\,2009ip or the pre-SN outburst of SN\,2006jc. The timescale is more reminiscent of the ``2012a phase'' of SN\,2009ip (Mauerhan et al. 2013a), and is also similar to that of SN\,2002bu (Smith et al. 2011) and SN\,1997bs (Van Dyk et al. 2000), both of which peaked at slightly fainter levels but had similar rise and decay times. 

The earliest available colour photometry of the precursor was obtained by {\it HST} during a relatively bright phase on 2001 July 7, at which point the source exhibited $B-I\approx0.7$\,mag. By 2005 March 10, after the source had dimmed by $>1$\,mag, the colour became substantially redder with $B-I\approx1.2$\,mag and $B-V\approx0.4$\,mag. 

Our first reliable colour measurement of the 2014 outburst was obtained only after the bright 2014b phase had begun. On June 7, we measured $B-I\approx0.9$ and $B-V\approx0.3$\,mag. By the outburst peak, near June 16 through July 2, the source had reddened only marginally to $B-I\approx1.0$ and $B-V\approx0.4$\,mag. Meanwhile, the extinction-corrected $UVW1-V$ colour of SN\,Hunt\,248, shown in Figure~\ref{fig:color}, began at a relatively red value of $\sim1.7$\,mag, and continually became bluer until reaching $\sim0.3$\,mag near the peak of 2014b. The $B-V$ colour evolution, on the other hand, is relatively flat and always bluer than $UVW1-V$, which suggests that the UV colour evolution is driven by an intrinsic spectral change in the \textit{UVW1} band (as opposed to variable extinction). By comparison, the Type IIn SN\,2011ht, which we will spectroscopically compare to SN\,Hunt\,248 in \S2.2, exhibits a rapid blueward shift of the $UVW1-V$ colour before peak, while its $B-V$ colour remains relatively flat like that of SN\,Hunt\,248. Clearly SN\,2011ht is intrinsically much more luminous in the UV than SN\,Hunt\,248, indicating a substantial difference in outflow energetics.

\begin{figure*}
\includegraphics[width=4.5in]{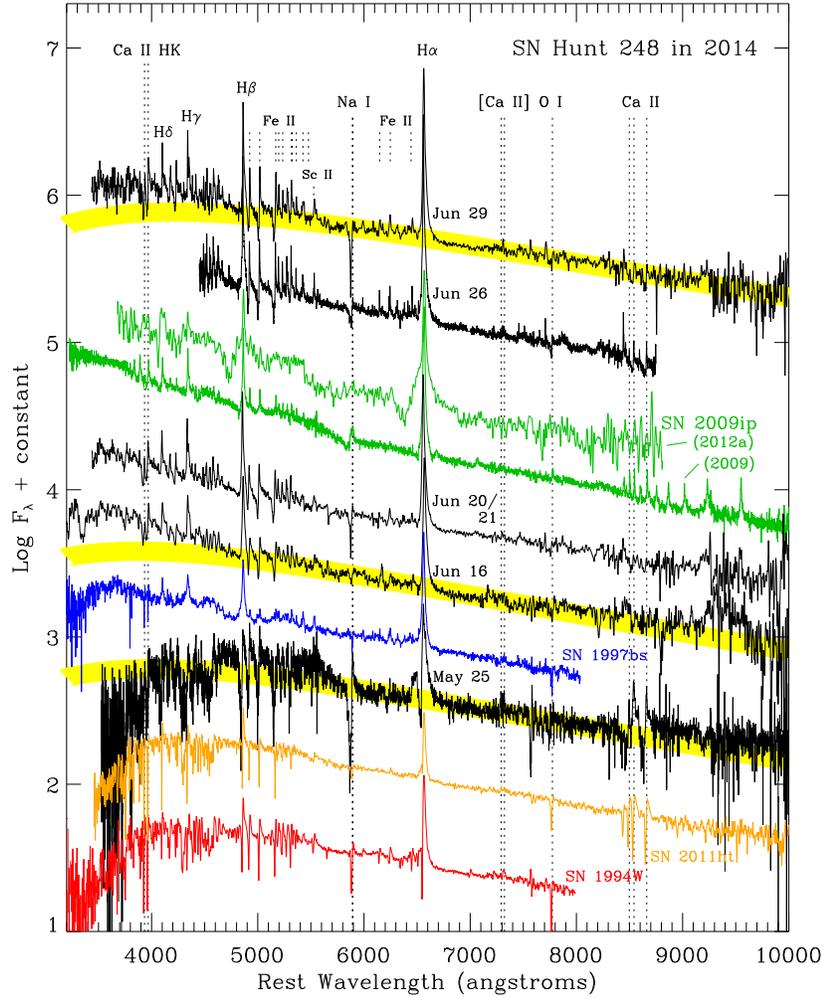}
\caption{Spectra of SN\,Hunt\,248 during the 2014a (May 25) and 2014b phases (June 16, 20/21 average, 26, and 29). Blackbody curves having $T_{\rm eff}=7200$\,K, 7500\,K, and  6900\,K are superposed on the spectra for May 25, June 16, and Jun 29 (yellow curves). All spectra have been corrected for extinction using $A_V=0.14$\,mag and assuming $R_V=3.1$. Note the strong blue continuum on June 16, near the peak of the 2014b phase. The following comparison spectra are included: Type IIn SN 1994W (red; Chugai et al. 2004), SN\,2011ht (Mauerhan et al. 2013b), SN\,2009ip during its discovery outburst in 2009 and in the 2012a phase of the supernova (green; Smith et al. 2010; Mauerhan et al. 2013a), and the SN impostor SN\,1997bs (blue; Van Dyk et al. 2000). All have been extinction corrected with adopted values from the literature. Prominent emission lines are marked. }
\label{fig:spec_wide}
\end{figure*}

\begin{table}\scriptsize\begin{center}\begin{minipage}[bp]{3.2in} \setlength{\tabcolsep}{3.3pt}
      \caption{Spectroscopic observations.}
\centering
  \begin{tabular}[bp]{@{}llcc@{}} 
  \hline
 Date                  & Instrument                 & $R\,(\lambda/\delta\lambda)$ & Wavelength range  \\                            
                            &                                     &                                                      & (\AA)  \\
\hline
\hline
2014-05-25       &    Keck/LRIS              &   800--1400    &   3500--10,000          \\ % (0568.6) 
2014-06-16       &    LCOGT/FLOYDS   &      300--700   &    3200--10,000                             \\ % (6175.6)   
2014-06-20       &    Lick/Kast   &      500--900       &    3300--10,400                              \\ % (6175.6)   
2014-06-21       &    Lick/Kast   &      500--900       &    3300--10,400                             \\ % (6175.6)   
2014-06-26       &    Keck/DEIMOS        &  1300--2700  &  4440--9660 \\ % (6178.1)     
2014-06-26       &    Keck/DEIMOS\,``hires''      &    4000--6000  &  5000--7500 \\ % (6178.1)     
2014-06-29       &    Lick/Kast   &      500--900       &    3300--10,400                       \\ % (6175.6)   

\hline
\end{tabular} \end{minipage} \end{center}
\end{table}

\subsection{Spectroscopy}

We obtained our first optical spectrum of SN\,Hunt\,248 on 2014 May 25, several days after its discovery. 
We used the Low Resolution Imaging Spectrometer (LRIS; Oke et al. 1995) on the Keck-I 10\,m telescope. For the blue channel, we employed the 600/4000 grism with a clear filter, covering a wavelength range of $\sim 3010$--5600\,{\AA} with a dispersion of $\sim 0.63$\,{\AA}\,pixel$^{-1}$, which yielded a FWHM resolution of $\sim 4$\,{\AA}. For the red channel we used the 400/8500 grating centered near 7809\,{\AA} with a clear filter, covering a wavelength range up to $\sim 9500$\,{\AA} with a dispersion of $\sim 1.16$\,{\AA}\,pixel$^{-1}$, giving a FWHM resolution of $\sim 6.9$\,{\AA}. Two integrations of 450\,s each were obtained
 at an airmass of 1.05, with the slit at the parallactic angle
(Filippenko 1982). Flat-field and wavelength calibration were performed 
using spectra of continuum and He-Ne-Ar emission sources internal to the instrument. 
Flux calibration was provided by observations of the standard stars Feige~34 and BD+26$^\circ$2606. 
Data reduction and calibration were performed using standard
 IRAF routines. For all spectra presented in this paper, the wavelength scale was corrected for the
redshift of NGC 5806 ($z = 0.0045$).

A spectrum was obtained on 2014 June 16 using LCOGT's  FLOYDS robotic spectrograph (Sand et al. 2011) at the Faulkes Telescope North. This observation provided our first epoch of the 2014b event, near peak brightness. FLOYDS uses a 235\,lines\,mm$^{-1}$ grating and cross-dispersing prism to provide a spectral resolving power ($R=\lambda/\delta\lambda$) approximately in the range 300 to 700 over a wavelength range of 3200--10,000\,{\AA}.

Low-resolution spectra were also obtained with the Kast spectrograph (Miller \& Stone 1993) on the 3\,m Shane reflector at Lick Observatory on 2014 June 20, 21, and 29 (each of these epochs occur during or after the 2014b peak). For the red and blue channels of the instrument, we observed with the 300/7500 grating and 600/4310 grism, respectively.  A D55 dichroic resulted in a crossover wavelength of $\sim 5500$\,\AA, and the total wavelength coverage was 3,300--10,400\,{\AA}.  Observations were 
performed through a 2{\arcsec} slit, and the FWHM spectral resolutions were $\sim 11$ and $\sim 6$\,\AA\ on the red and blue sides, respectively.

In addition, spectra were obtained on 2014 June 26 using the Keck-II 10\,m telescope and the Deep Imaging Multi-Object Spectrograph (DEIMOS;
Faber et al.\ 2003). The 600\,lines\,mm$^{-1}$ grating was used, providing a dispersion of 0.65\,{\AA}\, pixel$^{-1}$ through a 1{\arcsec}
slit, and a FWHM resolution of 3.5\,{\AA}. The wavelength coverage with this setting is 4440--9,660\,{\AA}.
The 1200\,lines\,mm$^{-1}$ ``hires'' grating was also utilized with the 0\farcs8 slit, which provided a dispersion of 0.33\,{\AA}\, pixel$^{-1}$, FWHM = 1.1--1.6\,{\AA}, and a wavelength range of 5000--7500\,{\AA}. Both settings included a GG455 order-blocking filter. Table\,4 is a summarized list of our spectroscopic observations.

\begin{figure}
\includegraphics[width=3.3in]{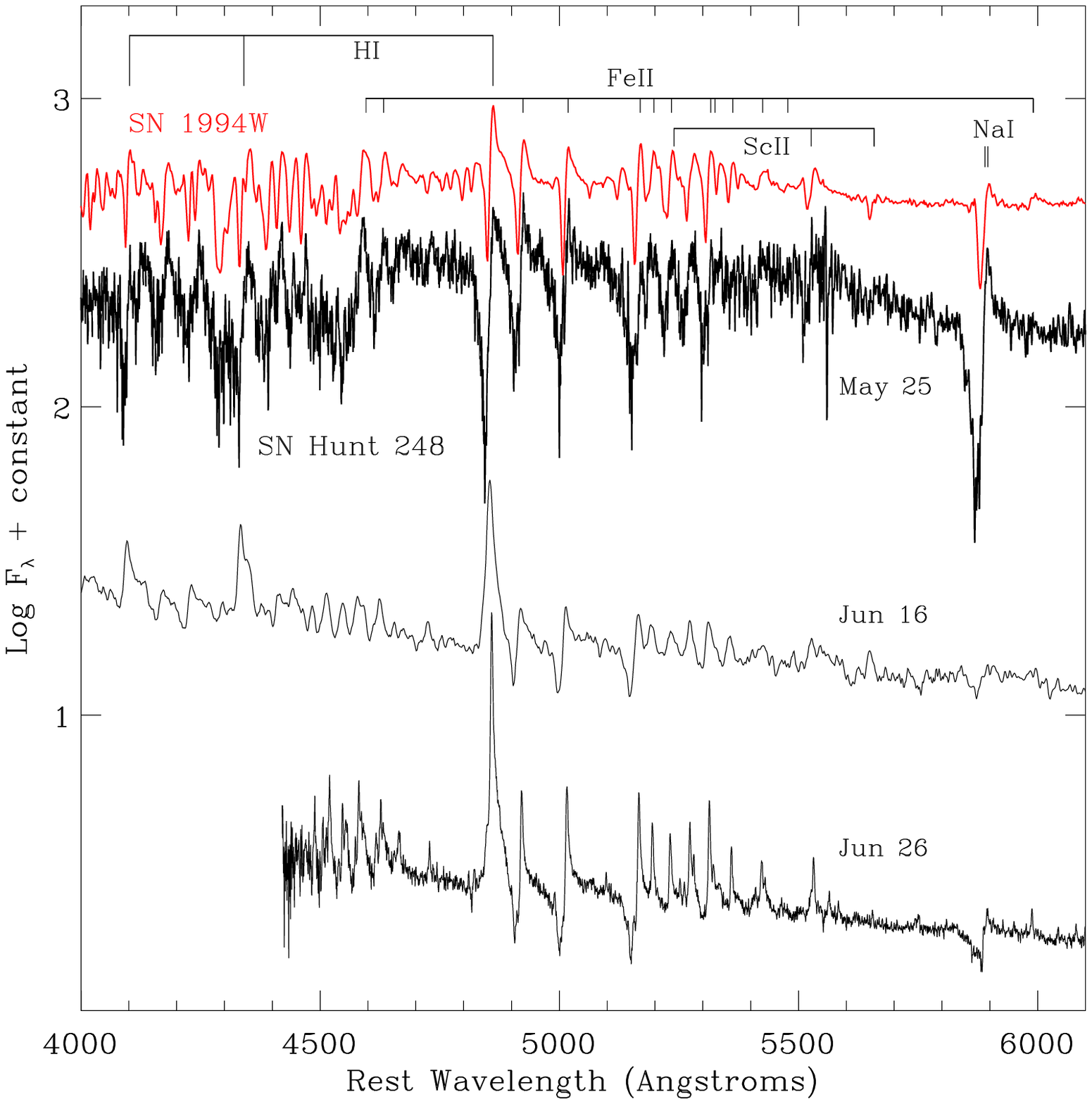}
\includegraphics[width=3.3in]{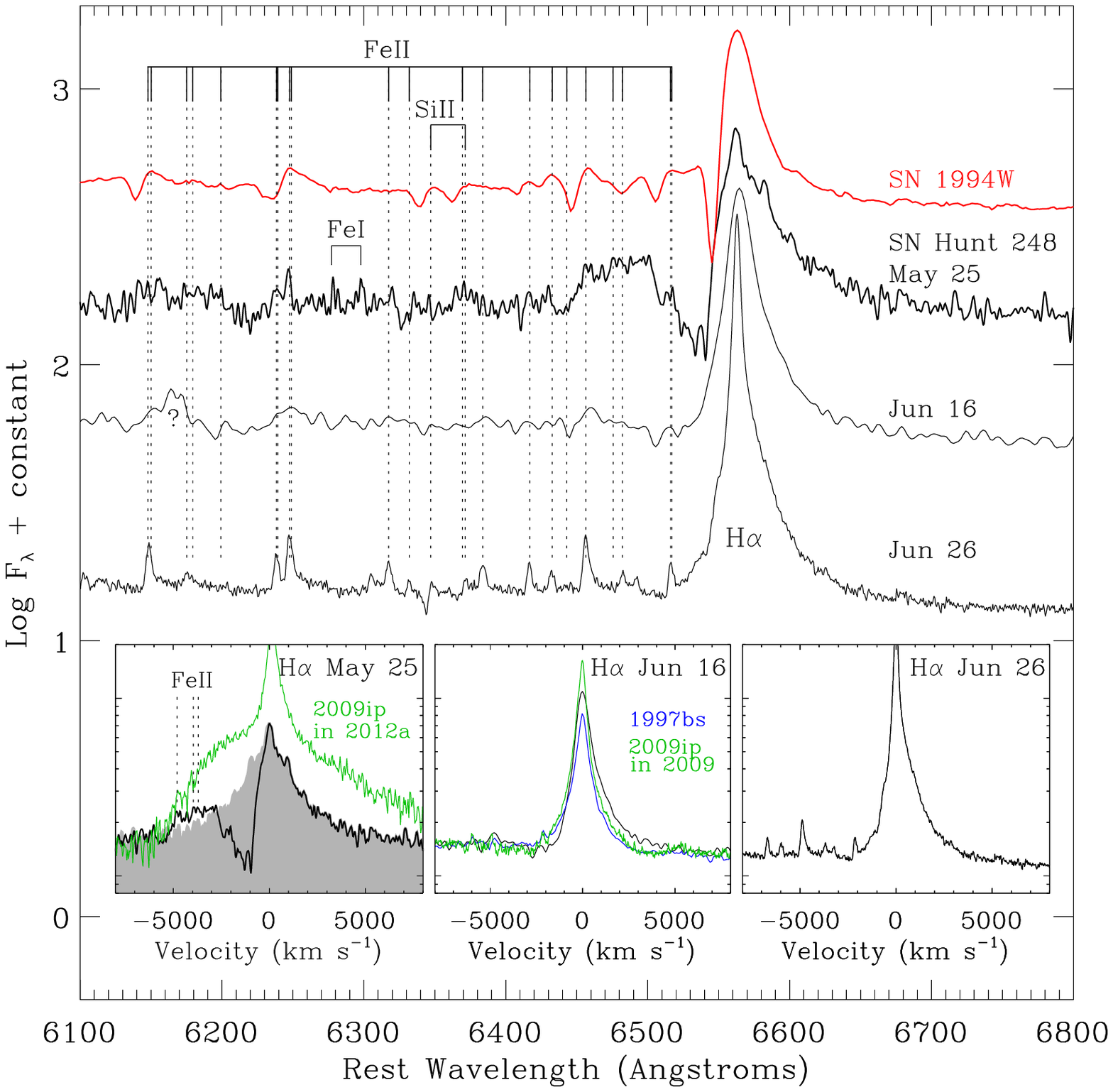}
\caption{Spectra of SN\,Hunt\,248 in the vicinity of 
H$\alpha$ (lower panel) and the rich Fe~{\sc ii} complex near 5000~{\AA} (upper panel). The inset frames are plotted with respect to the H$\alpha$ velocity. For the May 25 inset the grey background shows the mirror reflection of H$\alpha$ from the red into the blue. SN\,1994W (red; Chugai et al. 2004), the SN impostor SN\,1997bs (blue; Van Dyk et al. 2000), and SN\,2009ip (green) during its discovery outburst in 2009 (Smith et al. 2010) and during its 2012a SN phase (Mauerhan et al. 2013a) are also plotted for comparison.} 
\label{fig:spec_ha}
\end{figure}

\subsubsection{Spectral evolution}
The optical spectral sequence of SN\,Hunt\,248's 2014 evolution is shown in Figure~\ref{fig:spec_wide}. The spectra have been extinction corrected assuming $R_V=3.1$ (Cardelli, Clayton, \& Mathis 1989). On May 25, during the 2014a phase, the source exhibits a blackbody temperature of $T_{\rm eff} \approx 7200$~K and is dominated by H$\alpha$ P-Cygni and numerous transitions of Fe~{\sc{ii}}, most of which exhibit P-Cygni profiles. Na~{\sc i}~D P-Cygni (absorption dominated) and the Ca~{\sc ii} triplet at 7500--8000~{\AA} are also prominent features. Weak [Ca~{\sc ii}] emission at $\lambda\lambda$7291, 7323 and O~{\sc i} P-Cygni (absorption-dominated) at 7774~{\AA} are detected as well, in addition to weak P-Cygni features of Sc~{\sc ii} and Si~{\sc ii}, and weak emission from Fe~{\sc i}. Near the blue end of the spectrum, strong line blanketing from metals, including  Ca~{\sc ii} H\&K, suppresses the flux shortward of 5000~{\AA}. Overall, the 2014a spectrum exhibits the characteristics of a cool, dense outflow, very similar in appearance to the SNe~IIn-P 1994W and 2011ht during their cooler post-peak phases (Mauerhan et al. 2013b). However, the detection of [Ca~{\sc ii}], which normally de-excites back to the ground state in a cool, dense gas such as this, implies the presence of an additional low-density component in the flow.

The line profiles of SN\,Hunt\,248 are shown more clearly in Figure~\ref{fig:spec_ha}. On May 25, H$\alpha$ exhibits a line core having a FWHM velocity of $\sim 1000$\,km\,s$^{-1}$, while the bases of the lines exhibit broad Lorentzian wings that extend out to $\sim 3000$\,km\,s$^{-1}$. On the blue side of the H$\alpha$ line, the superposition of the extended Lorentzian wing and multiple Fe~{\sc ii} transitions might (at first glance) give the false appearance of a high-velocity edge to the broad H$\alpha$ emission. The flux minima of the P-Cygni absorption components indicate an average outflow velocity of $\sim 1200$\,km\,s$^{-1}$ during 2014a. The blue edges of the absorption components, particularly Na~{\sc i}~D, extend to $\sim 2600$\,km\,s$^{-1}$, indicating the presence of faster material.

The spectrum of SN\,Hunt\,248 near the peak of the 2014b phase on June 16 exhibited some substantial changes from the 2014a phase. Balmer emission lines became dominant, while Na~{\sc i}, Fe~{\sc ii}, and the metal line blanketing features at the blue end of the spectrum weakened, giving rise to a stronger blue continuum. This implies that line blanketing must have also significantly decreased in the UV, which could explain the $UVW1-V$ colour evolution shown in Figure~\ref{fig:color}. But the overall slope of the spectrum at wavelengths $>5000$~{\AA} has not changed significantly from the earlier spectrum on May 25, and it is consistent with a blackbody temperature of $\sim 7500$~K, only a little hotter than the 2014a phase. The Fe~{\sc ii} lines near 5000~{\AA} are still evident as P-Cygni profiles, though the Balmer P-Cygni absorption components seen earlier are no longer obvious. But Lorentzian wings extending to $\pm3000$\,km\,s$^{-1}$ remain present in the Balmer profiles. At this stage, SN\,Hunt\,248 closely resembles the impostor SN\,1997bs, having a very similar continuum slope and a Balmer-dominated emission spectrum. The 2014b spectrum also shares spectral similarities with SN\,2009ip during its discovery eruption in 2009 (Smith et al. 2010), but SN\,Hunt\,248 is not quite as blue in the continuum and not as high in ionisation temperature (e.g., SN\,2009ip exhibited substantial emission from He {\sc i)}. The H$\alpha$ profile during the 2014b phase also appears similar to that of SN\,1997bs and SN\,2009ip, having comparable core FWHM values and similarly broad Lorentzian wings. The velocity implied by SN\,Hunt\,248's H$\alpha$ profile, however, is not nearly as fast as that of SN\,2009ip during its 2012a event (i.e., the onset of SN\,2009ip's SN explosion; Mauerhan et al. 2013a).

By June 26, the Balmer lines increased further in strength relative to the continuum, but this time the Fe~{\sc ii} P-Cygni lines near 5000~{\AA} and Na~{\sc i} have become stronger as well. Narrow emission features of Fe~{\sc ii} blueward of H$\alpha$ also become prominent, though without obvious P-Cygni profiles. At first glance, the blue edge of the Na~{\sc i}~D line appears to extend to higher velocity at this stage ($>3000$\,km\,s$^{-1}$) than during 2014a, but it is likely that He~{\sc i} is influencing the line profile; in fact, the absorption component of the P-Cygni line near Na~{\sc i}/He~{\sc i} appears to have a double trough, whose two minima are separated by $\sim 20$~{\AA}, which is consistent with the wavelength separation between Na~{\sc i} and He~{\sc i}. The H$\alpha$ profile on June 26 developed an asymmetric shape with a red shoulder, but does not redevelop a P-Cygni absorption component, as the Fe~{\sc ii} transitions do. Our most recent spectrum on June 29 appears very similar to that on June 26, although the more extended short-wavelength coverage on this date shows that the blue continuum has weakened since June 21 and the effective temperature of the source has become slightly cooler.  

\section{Discussion}

If SN\,Hunt\,248 were a bona fide SN explosion, it would be spectroscopically classified as a Type IIn, based on the presence of namesake ``narrow'' emission lines (Schlegel 1990; Filippenko 1997). However, the collective characteristics of this object, including its comparatively low peak luminosity, its slow outflow velocities, and relatively weak UV continuum, suggest that the 2014 events were substantially less energetic than what we would normally expect from a core-collapse SN. It thus appears very likely that SN\,Hunt\,248 is a SN impostor --- that is, a nonterminal outburst from a massive star in a post-main-sequence phase of its evolution. Indeed, the $\sim 1$\,mag variability of the stellar precursor observed from 1997 through 2005 is also consistent with this interpretation, as such behaviour is reminiscent of the S~Doradus brightness variations exhibited by LBVs and their kin.  Below, we explore the nature of the culprit star and the physical processes that could have led to the recent outbursts.

\begin{figure}
\includegraphics[width=3.4in]{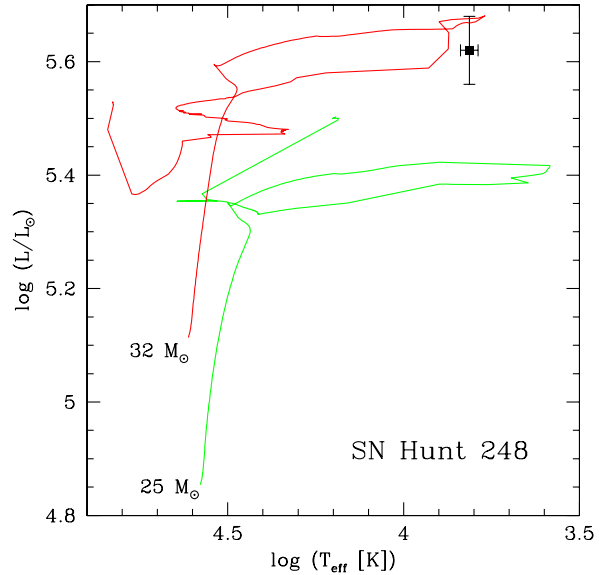}
\caption{The precursor of SN\,Hunt\,248 in the HR diagram. The stellar parameters were calculated using the ``quiescent'' {\it HST} photometry from 2005 (see Table 2). The estimate assumes the emission is purely stellar and should be interpreted with caution (see text). The isochrones are from the Geneva rotating stellar models (Ekstr{\"o}m et al. 2012).} 
\label{fig:hr}
\end{figure}

\subsection{The nature of the precursor star}
\subsubsection{HR diagram and evolutionary status}
To estimate the physical parameters of the precursor star of SN\,Hunt\,248, we used the {\it HST} photometry from the ``quiescent'' epoch on 2005 Mar. 10 to place the star on the Hertzsprung-Russell (HR) diagram (correcting for \textit{Galactic} foreground extinction only, and assuming a single star). The results indicate a stellar luminosity of $L=(4\pm1)\times10^{5}\,{\rm L}_{\odot}$ (uncertainty dominated by distance) and an effective temperature of $T_{\rm eff}=6500\pm300$\,K, which corresponds to a blackbody radius of $3.5\times10^{13}$~cm ($500\,{\rm R}_{\odot}$). We estimated this temperature by comparing the reddened $B-V$ colours to Castelli \& Kurucz (2003) supergiant model atmosphere spectra, using {\sc synphot} in the {\sc iraf/stsdas} package. Models of 6750\,K and 6250\,K did not produce satisfactory fits, which is how the uncertainty of 300\,K was estimated. No combination of the stellar models (nor pure blackbody curves) and reddening would simultaneously fit both the $B-V$ and $V-I$ colours, which we suspect is attributable to intrinsic excess emission in the $I$ band. The $I$-band data point exhibits an excess of $\sim0.3$\,mag over our best fit to the $B-V$ colours. 

  \begin{figure*}
\includegraphics[width=5in]{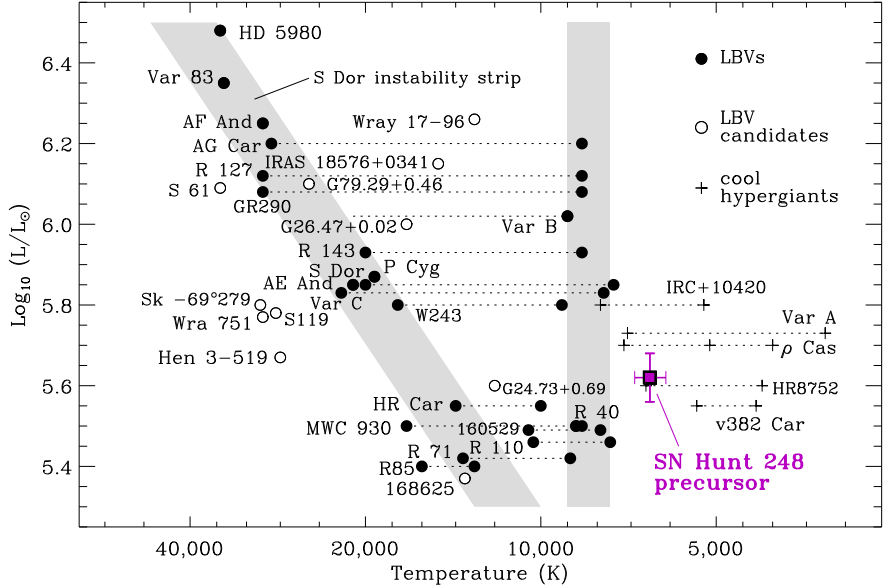}
\caption{Reproduction of Figure 1 from Smith et al. (2004), but with updated values from Smith \& Tombleson (2014), showing the HR diagram for LBVs and related stars, including SN\,Hunt\,248 (purple square). The diagonal and vertical grey strips illustrate the regions of the S~Doradus instability strip and the minimum temperature strip for classical LBVs near visual maximum, respectively; the white region between them is known as the ``yellow void." The stellar precursor of SN\,Hunt\,248 lies in the region occupied by cool hypergiants, such as $\rho$ Cas (although we have not taken into account the possibility of additional circumstellar extinction).} 
\label{fig:hr2}
\end{figure*}

Comparison of $L$ and $T_{\rm eff}$ for SN\,Hunt\,248 to Geneva rotating stellar models (Ekstr{\"o}m et al. 2012) is illustrated in Figure~\ref{fig:hr}. The results suggests an initial stellar mass of $\sim 30\pm2\,{\rm M}_{\odot}$ for the precursor of SN\,Hunt\,248. This value uncertainty was estimated by eye from a visual comparison of the position of SN\,Hunt\,248 in Figure~\ref{fig:hr} with respect to the red termini of the 25\,M$_{\odot}$ and 32\,M$_{\odot}$ models. However, our mass estimate should be taken with caution, as our analysis was performed under the assumption that the extinction-corrected (Galactic only) colours represent the true colour of the stellar photosphere. We have not accounted for the possibility of additional reddening from material local to the source. To investigate the variations in extinction in the vicinity of the source, we examined the colour distribution of neighboring stars in the {\it HST} images, within $\sim100$ pc, in projection, of SN\,Hunt\,248; however, the results were not very illuminating, as the sample of stars from which to draw this information were few and faint, and exhibited a very large scatter in $B-V$ and $V-I$ colour space. Still, there is spectral evidence that the extinction from host-galaxy ISM along the line of sight to the source is very low, which we present later.  Erstwhile extinction from pre-existing dusty CSM that was only present during the precursor phases (before the recent UV-bright eruptions), although purely speculative, is another potential source of error of which we should be cautious, for reasons described below.
  
Figure~\ref{fig:hr2} is a reproduction from Smith et al. (2004; their Figure~1) showing the distribution of LBVs and related stars in the HR diagram. The precursor of SN\,Hunt\,248 lies within a region of $L$--$T_{\rm eff}$ space where the cool hypergiants reside, such as $\rho$ Cas, IRC~$+$10420, HR~8752, and Var A in M33.
It is thus plausible that SN\,Hunt\,248 was in a cool hypergiant phase during the decade before its 2014 eruption. Our estimated initial mass is also consistent with this interpretation, as it is very similar to the inferred initial mass of the cool hypergiant $\rho$ Cas (30\,${\rm M}_{\odot}$; de Jager 1998). 

Cool hypergiants produce dense, opaque winds (Davidson 1987), and this could explain the colour evolution of the precursor to SN\,Hunt\,248. For example, as the source decreased in brightness by $\Delta I \approx 1.1$\,mag between its relatively bright phase in 2001 and its ``quiescent'' state in 2005, it exhibited an increasingly red colour, from $B-I\approx0.7$ to 1.2\,mag. Classical LBVs and supergiants, on the other hand, generally get bluer as they fade, not redder --- but in these cases the decreasing brightness is usually associated with the physical contraction of the stellar radius and an increase in the effective temperature. Thus, the reddening of the star as it dimmed could actually be the signature of a dense, opaque wind in the process of formation between 2001 and 2005. 

On a related note, the condensation of dust in the outer wind or in shells of CSM could also have played an important role in the colour evolution. Indeed, dusty CSM shells have been observed around cool hypergiants, including $\rho$~Cas (Sargent 1961), and are also a common feature of LBVs, such as $\eta$ Car, AG Car, HR Car, P Cygni, R 127, and candidate LBVs (cLBV) such as HD\,168625 (O'Hara et al. 2003) and the large sample of cLBVs surrounded by mid-infrared shells detected with the MIPS instrument aboard the \textit{Spitzer Space Telescope} (Wachter et al. 2010). In the case of SN\,Hunt\,248, \textit{Spitzer} Warm Mission data at 3.6\,${\micron}$ and 4.5\,${\micron}$ wavelengths were obtained for the host galaxy NGC 5806, and we have analysed these data to constrain the infrared flux density to $<24.8\,\mu$Jy and $<35.0\,\mu$Jy, respectively. Unfortunately, at the distance of NGC~5806, the limits at these wavelengths do not place meaningful constraints on a potential thermal dust source. For example, at the distance of NGC 5806, a dusty CSM shell similar to that around the LBV HD168625 ($F_{4.7\tiny{{\micron}}}=0.6$\,Jy at 1200\,pc; Robberto \& Herbst 1998) would exhibit a flux of $\sim 1.5\times10^{-9}$\,Jy in a comparable wavelength domain, four orders of magnitude fainter than our \textit{Spitzer} limits. Even a very dusty IR-luminous hypergiant such as NML Cygni (Schuster et al. 2009), if placed at the distance of NGC\,5806, would exhibit fluxes less than 0.1 times our limits. 

\begin{figure}
\includegraphics[width=3.3in]{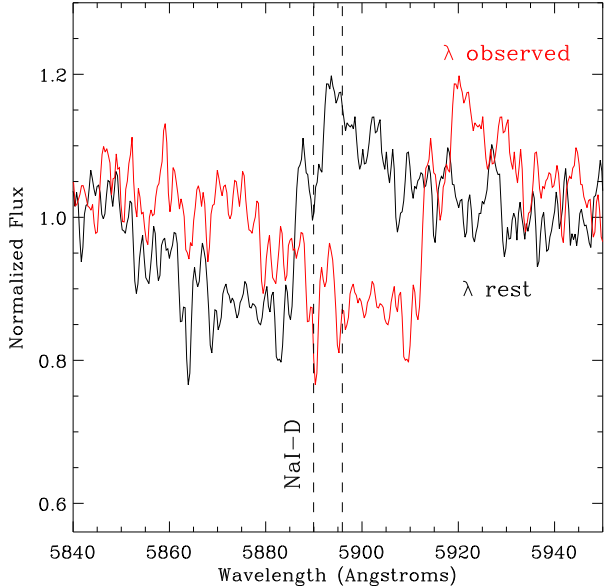}
\caption{Keck/DEIMOS spectrum of the Na\,{\sc i} doublet region, showing the broad feature from the outflow in SN\,Hunt\,248, and the subtle narrow absorption features from the ISM (where the dashed lines cross the spectra). The red spectrum exhibits the \textit{observed} wavelength scale, where the dashed lines cross the Galactic absorption components. The black spectrum has been corrected for a redshift of 0.0045, where the dashed lines cross the host's absorption components. 
The narrow absorption is weak in both cases, and suggests a low value of interstellar extinction.} 
\label{fig:naD}
\end{figure}

The amount of extinction from the host ISM can be estimated by examining the strength of the narrow Na\,{\sc i} doublet absorption feature at the redshift of NGC 5806. Figure\,\ref{fig:naD} shows this region of the spectrum from our highest spectral resolution data obtained with Keck/DEIMOS on June 26. Both the Galactic and host-galaxy absorption profiles can be seen, but the weakness of these features results in a low signal-to-noise ratio. Both components of the doublet are discernible for the Galaxy, but only the stronger shorter-wavelength feature of the doublet ($D_2$ at $\lambda5889.95$\,{\AA}) can be distinguished clearly at the redshift of NGC 5806. The equivalent width (EW) of this feature is $0.2\pm0.1$\,{\AA} for both the Galaxy and for NGC~5806. According to Poznanski, Prochaska, \& Bloom (2012), this corresponds to a small colour excess of $E(B-V) \approx 0.03\pm0.02$\,mag for each, or $0.09\pm0.03$\,mag total, corresponding to $A_V=0.19\pm0.07$\,mag, assuming an $R_V=3.1$ extinction relation (Cardelli, Clayton, \& Mathis (1989). This is consistent with our adopted $A_V=0.14$\,mag for the Galaxy, suggesting a comparably small amount of extinction in the host. Therefore, additional absorption from the host ISM is not likely to have a significant impact on our photometric analysis, or on interpretation of the precursor star's evolutionary state. 

There could, however, have been additional reddening from CSM dust back in 2005 that has since been destroyed by UV photons (e.g., Wesson et al. 2010), either from the recent 2014a event or earlier outbursts between 2005 and 2014 that were not observed. If so, we would not see the absorbing effect of this dust in the more recent spectra of SN\,Hunt\,248, and we would thus have no way of accounting for it in our photometric analysis of the stellar precursor. However, we can consider the potential influence of CSM dust by asking what amount of additional reddening, if accounted for, would cause SN\,Hunt\,248's location in the HR diagram to be moved out of the cool-hypergiant section --- that is, moved hotter than 8500\,K. According to Flower (1996), the intrinsic colour of a supergiant with, for example, $T_{\rm eff}=8721$\,K is $(B-V)_0\approx0.08$\,mag. Hypothetically, if SN\,Hunt\,248 were this hot in 2005 and had this same intrinsic colour, then the observed colour of $B-V=0.39$\,mag would imply a total extinction of $E(B-V)\ge0.31$\,mag, or $A_V=0.96$\,mag, assuming $R=3.1$. Subtracting the Galactic foreground value would then imply $A_V=0.82$\,mag of unaccounted extinction. In this case, the inferred luminosity would be a factor of $\sim 2$ higher, or $\sim 9\times10^5\,{\rm L}_{\odot}$. This would place the precursor of SN\,Hunt\,248 near R143 or S Dor during their cool states (see Figure~\ref{fig:hr2}). We remind the reader, however, that there is no evidence for this hypothetical amount of additional circumstellar extinction in any of the available data. We thus move forward in favour of the cool-hypergiant classification for the precursor star in 2005.

\subsection{The nature of the 2014 outburst(s)} 
\subsubsection{The 2014a phase: a super-Eddington outflow}
In its quiescent state during 2005, the stellar precursor's luminosity and estimated initial mass indicate that the electron-scattering Eddington ratio at that time was $\Gamma = (\kappa_e L)/4 \pi GMc \approx0.3$. Note, however, that the actual stellar mass for this eruptive variable star has probably decreased substantially from its initial value, as a result of mass loss via winds and, perhaps, former eruptions over the course of its post-main-sequence life.  For a significantly lower mass of, say, $M=11\,{\rm M}_{\odot}$ (near the lower mass end for stars that become cool hypergiants with opaque winds; Smith et al. 2004), the Eddington ratio would have been $\Gamma\approx0.9$, very near the tipping point for the star. Nine years later, at the onset of the 2014a outburst, the increase in luminosity implies that the Eddington ratio had already climbed to a value in the range $\Gamma\approx4$--12 (for the same conservative range of stellar masses used above, 11--32\,M$_{\odot}$). It is thus very plausible that the 2014a eruption resulted from the star crossing over the Eddington limit, with which it appears to have been flirting during 1997--2005. A super-Eddington wind with $\Gamma>10$ can drive strong mass loss on the order of $\sim 0.1\,{\rm M}_{\odot}$\,yr$^{-1}$ (Owocki et al. 2004).

The average outflow velocity of 1200\,km\,s$^{-1 }$ during the 2014a phase is several times faster than LBV winds and an order of magnitude faster than the winds of cool hypergiants, but is in the observed range for SN impostors (Smith et al. 2011). This velocity, in addition to the range of possible Eddington ratios ($\Gamma=4$ to 12), suggests that the 2014a event could have been explosive, as opposed to wind-like. Consistently, the spectral evidence for fast outflowing material having $v \approx 2600$\,km\,s$^{-1}$, as indicated by the blue edges of the P-Cygni absorption components, suggests that there could be a fast blast wave of tenuous matter at radii farther out in the flow; this could be where the [Ca~{\sc ii}] emission originates (see Figure~\ref{fig:spec_wide}). 

\subsubsection{The 2014b phase -- CSM interaction}
The rapid onset of the 2014b phase, which followed a rather slow, modest increase in brightness during 2014a, implies that a relatively sudden event took place. CSM interaction is an attractive possibility that could explain both the quick luminosity jump and the subsequent strengthening of the blue continuum and Balmer lines in the spectrum near peak on June 16. SN\,Hunt\,248 during 2014b could thus be similar to a SN IIn, albeit with a much lower energy shock (e.g., Smith 2013). Indeed, most SNe~IIn, such as SN\,2011ht, begin hot and UV bright and \textit{later} become cool and dense, with the UV flux declining ahead of the optical. SN\,Hunt\,248 appears to evolve in the other direction, beginning cool and dense and subsequently peaking late in the UV. Then again, SN\,Hunt\,248 was discovered in an early phase several weeks before the 2014b phase began and climbed to peak brightness. Had we taken our first spectrum nearer to the 2014b peak, there would have been no clear indication of the initial cool, dense phase. Still, the UV evolution relative to the optical, illustrated in Figure~\ref{fig:color}, would still have appeared noticeably different than a SN IIn.

LBV-like stars and cool hypergiants are known to have extended winds and are commonly surrounded by circumstellar nebulae. CSM interaction in the case of SN\,Hunt\,248 would thus be unsurprising if a blast wave was generated by the giant outburst. If interaction began in 2014b, when the 2014a outflow encountered an outer wind or a CSM shell that was moving at a substantially lower velocity, we might have expected to witness a change in the velocities associated with the P-Cygni absorption minima (e.g., in the Fe~{\sc ii} lines), once the shock hit the slower outer wind and illuminated the pre-shock gas. No such velocity shift was seen in the spectra between May 25 (2014a phase) and June 16 (peak of 2014b phase). However, in the absence of a strong luminous shock like a typical SN IIn, there may have been insufficient flux of UV/X-ray photons to illuminate the outer wind and imprint its emission-line signature on the source's spectrum (Chevalier \& Fransson 1994). Additionally, if the CSM geometry was aspherical and different from that of the impinging outflow, then the spectral features present during 2014a could remain more or less unchanged during the 2014b interaction, aside from dilution by the added continuum from CSM-interaction luminosity.

Alternatively, the 2014b phase could have begun as material from a second outburst plowed through slower dense material from the 2014a outflow. In this case, the observational signature of the inner fast component would be obscured by the outer envelope from 2014a, the kinematics of which would determine the morphology of spectral line profiles during both stages. Eventually, if the shock is strong enough, we might expect to see a velocity transition once it overruns the edge of the 2014a outflow and encounters the wind/CSM from an earlier phase (assuming that the 2014a outflow was from a discrete eruption and not a more slowly developing super-Eddington wind). Such a transition might not be detectable, however, if the shock is severely weakened by then, and/or if the outer CSM does not have sufficient density to create an observable interaction. 

In either of the two CSM-interaction scenarios considered above, the majority of the 2014b luminosity is generated by the conversion of kinetic energy into radiation. Between the onset of the 2014b phase through  July 12 (the end of the coverage presented here), the total radiated energy of the source (without bolometric correction) is $L_{\rm rad} \sim 10^{49}$\,erg. The efficiency of the conversion of kinetic energy into radiation is not reliably inferable from the available data. However, if we were to hypothetically assume an efficiency factor of 0.1, then this would imply a kinetic energy of $\sim10^{50}$\,erg for the 2014 eruption. Thus, for an average outflow velocity on the order of 1000\,km\,s$^{-1}$, the implied ejected mass is on the order of $\sim 1\,{\rm M}_{\odot}$. 

\subsection{The cool hypergiant -- LBV connection}
The likely conjecture that SN\,Hunt\,248 was in a cool hypergiant phase before undergoing a giant outburst has interesting and important consequences for stellar evolution, for  reasons described as follows. The HR diagram in Figure~\ref{fig:hr2} shows the location of the so-called S Doradus instability strip, a diagonal stretch of  $L$--$T_{\rm eff}$ space that is populated by LBVs in their hotter states of relative quiescence, near their visual minima. Within a particular region on this strip there is a dearth of LBVs having effective temperatures in the range of 15,000\,K to 21,000\,K and luminosities in the range of log~$L/{{\rm L}_{\odot}}<5.7\pm0.1$. Coincidentally, this is the same range of luminosities exhibited by the cool hypergiants that lie directly to the right (at cooler temperatures) of this vacant region of the HR diagram. 

Smith et al. (2004) proposed that the deficit of LBVs in this section of the S Doradus instability strip could be the result of the so-called ``bistability jump'' (Lamers et al. 1995; Vink et al. 1999; Vink \& de Koter 2002). Cooler than $T_{\rm eff} =$  21,000\,K (spectral type B1) there is a sudden ionisation-dependent jump in the line-driving parameters within the stellar wind, resulting in an abrupt increase in mass-loss rate and a substantial drop in the terminal velocity of the wind ($v_{\infty}$) with respect to the escape speed ($v_{\rm esc}$). This leads to a catastrophic increase in wind density. As a result, an opaque pseudo-photosphere develops in the wind, which presumably can cause LBVs that would otherwise lie on the S Doradus instability strip to shift cooler than 8500\,K in the HR diagram, into the cool hypergiant realm. Smith et al. (2004) suggested this to be the evolutionary path for $\rho$~Cas and the other cool hypergiants mentioned above ---  i.e., they \textit{are} the missing hot LBVs from the S Doradus strip, masquerading as cool supergiants by means of their dense, opaque winds. 

If the proposed connection between cool hypergiants and LBV phenomena described above is correct, then we might expect cool hypergiants to be capable of generating violent super-Eddington outbursts, like LBVs do. The 2014 eruption of SN\,Hunt\,248 is thus extremely interesting in this regard, providing the first observational example to solidify this connection. Furthermore, the case of SN\,Hunt\,248 demonstrates not only that cool hypergiants can experience giant outbursts, but that they can rival in peak luminosity the giant eruptions from stellar behemoths like $\eta$ Car, even though they stem from stars of relatively low initial mass. This achievement could have been made possible by the boost in luminosity provided by CSM interaction.   

\section{Concluding Remarks}
The case of SN\,Hunt\,248 provides a valuable new addition to the small sample of SN impostors for which we have information on the precursor. The 2014 eruption appears to have been the result of a super-Eddington outburst which experienced a subsequent phase of CSM interaction that resulted in a large boost in luminosity. SN\,Hunt\,248 provides the first observational connection between cool hypergiant stars and explosive LBV phenomena, lending credence to the hypothesis that cool hypergiants are the missing LBVs from the S~Doradus strip, hidden behind cool, dense, opaque envelopes (Smith et al. 2004). 

The looming fact remains that there is still no satisfactory explanation for what drives massive stars to generate super-Eddington outbursts or other explosive processes that produce luminous transient events. Multiple possibilities have been outlined by Smith et al. (2011) and Smith \& Arnett (2014), some of which might be applicable to SN\,Hunt\,248. These include runaway stellar pulsations, geyser-like eruptions caused by large opacity variations during post-main-sequence evolution, turbulent and explosive nuclear shell-burning, the pair-instability mechanism (probably only important for extremely massive stars), electron-capture SNe (probably important only for 8--$10\,{\rm M}_{\odot}$ stars), or failed SNe. Close stellar encounters in binary systems are another potential pathway for giant outbursts (e.g., see Smith 2011). This is an interesting possibility in the case of SN\,Hunt\,248, as nonconservative mass transfer in a massive binary could have led to the formation of the extended cool hypergiant envelope (e.g., see Chesneau et al. 2014), preceding the recent eruption. 

A more complete sample of SN impostors and higher cadence photometry and spectroscopy during multiple phases of their evolution will help elucidate the nature of these outbursts. The near-future development of deep high-cadence sky surveys such as the Large Synoptic Survey Telescope will undoubtedly lead to a significant improvement in our understanding of these objects. In addition, a more thorough investigation of massive stars in the Galaxy having resolved circumstellar nebulae that resulted from giant outbursts will help connect the observables of extragalactic transients to the physical properties and energetics of the outflows they produce. Once accompanied by a firmer theoretical foundation regarding the late stages of stellar evolution, the myriad of potential physical processes involved in generating SN impostors and other luminous transients may begin to be narrowed down with some confidence.

\section*{Acknowledgments}

\scriptsize 

We thank the anonymous referee for a detailed review, which
led to an improved paper. This work is based in part on observations made with the NASA/ESA {\it Hubble Space Telescope}, obtained from the Data Archive at the Space Telescope Science Institute, which is operated by the Association of Universities for Research in Astronomy, Inc., under NASA contract NAS 5-26555. This work is also based on observations made with the {\it Spitzer Space Telescope} and made use of the NASA/ IPAC Infrared Science Archive, which is operated by the Jet Propulsion Laboratory, California Institute of Technology under a contract with NASA. Some of the data presented herein were obtained at the
W.~M.\ Keck Observatory, which is operated as a scientific partnership
among the California Institute of Technology, the University of
California, and NASA; the observatory was made possible by the
generous financial support of the W.~M.\ Keck Foundation.  We are
grateful to the staffs at the Lick and Keck Observatories for their
assistance.  The supernova research of A.V.F.'s group
at U.C. Berkeley is supported by Gary \& Cynthia Bengier, the Richard
\& Rhoda Goldman Fund, the Christopher R. Redlich Fund, and the TABASGO
Foundation. KAIT and its ongoing operation were made possible by donations from Sun
Microsystems, Inc., the Hewlett-Packard Company, AutoScope
Corporation, Lick Observatory, the NSF, the University of California,
the Sylvia \& Jim Katzman Foundation and the TABASGO Foundation.  
This work is based in part on observations from the LCOGT network.
We have made use of the NASA/IPAC Extragalactic Database (NED), which is operated by the Jet Propulsion Laboratory, California Institute of Technology, under contract with NASA.

\end{document}